\input harvmac
\def\half{{1 \over 2}}

\def\>{{\rangle}}
\def\<{{\langle}}

\def\p{{\partial}}

\def\s{{\sigma}}

\def\L{{\Lambda}}
\def\l{{\lambda}}

\def\a {{\alpha}}
\def\b {{\beta}}

\def\ad {{\dot \a}}
\def\bd {{\dot \b}}

\def\d {{\delta}}

\def\t {{\theta}}

\def \t {{\theta}}

\Title{\vbox{\hbox{IFT-P.040/97}}}
{\vbox{\centerline{\bf An Introduction to Superstring Theory}
\centerline{\bf and its Duality Symmetries}}}
\bigskip\centerline{Nathan Berkovits}
\bigskip\centerline{Instituto
de F\'{\i}sica Te\'orica, Univ. Estadual Paulista}
\centerline{Rua Pamplona 145, S\~ao Paulo, SP 01405-900, BRASIL}
\bigskip\centerline{e-mail: nberkovi@ift.unesp.br}
\vskip .2in
In these proceedings for the First School on Field Theory
and Gravitation (Vit\'oria, Brasil), a brief introduction is given to
superstring theory and its duality symmetries. This introduction is
intended for beginning graduate students with no prior knowledge of
string theory. 

\Date{July 1997}
\newsec {Introduction and Motivation}

These proceedings are based on two lectures presented in April 1997 at the
First School on Field Theory
and Gravitation in Vit\'oria, Brasil. Instead of giving references in
these proceedings, I will refer the reader to the following review articles
which appeared recently on the hep-th bulletin board:
hep-th 9612254
(a review of
perturbative string theory by H. Ooguri), hep-th 9702201 (an introduction
to string dualities by C. Vafa), hep-th 9607021 (lectures on
superstring and M-theory dualities by J. Schwarz),
hep-th 9612121 (four lectures on
M-theory by P. Townsend), hep-th 9609176 (a lecture on T and S-dualities
by A. Sen), hep-th 9611050 (an introduction to $D$-branes by J. Polchinski),
and hep-th 9611203 (a review of supermembranes by M. Duff).
There are also two colloquia for a general audience which are
available on the hep-th bulletin board: hep-th 9607067 by J. Schwarz and
hep-th 9607050 by J. Polchinski. 

There are various motivations for studying superstring theory, both
mathematical and physical. Since I am a physicist, I will only
mention the physical motivations. When string 
theory was discovered in the early 1970's, 
it was origninally intended to be a model
for describing strong interactions. The basic discovery was 
that by extending the pointlike nature of particles
to one-dimensional extended 
objects
called strings, one could obtain S-matrix scattering amplitudes
for the fundamental particles   
which contained many of the properties found in scattering
experiments of mesons. As will be discussed in section 2, the
action for string theory is proportional to the area 
of a two-dimensional worldsheet, as
opposed to the action for point-particles which is based on
the length of a one-dimensional worldline. 

Amazingly, the masses and coupling constants of the fundamental
particles in string theory
are not inputs in the theory, but are instead fixed 
by consistency requirements such as Lorentz invariance and
unitarity. In fact, unlike theories based on point particles,
string theory
not only predicts the masses of the fundamental particles, but also
predicts the dimension of spacetime. In the simplest string theory, this
dimension
turns out to be 26, rather than the experimentally observed
spacetime dimension of 4. However, as will be discussed in section 5,
it is possible to `compactify' all but four of the dimensions to
small circles, in which case only four-dimensional spacetime is 
observable at low energies. 

For open string theory (where particles are represented
by one-dimensional objects with two ends), 
the particle spectrum contains a massless `gluon', as well as an infinite
number of massive particles whose masses and spins
sit on `Reggae trajectories'. 
These Reggae trajectories of massive particles
are welcome for describing strong
interactions since they are needed for producing scattering amplitudes
with the properties seen in experiments.  
Unfortunately, 
string theory also predicts fundamental particles which 
are not needed for describing strong interactions.
One of these particles is tachyonic, 
i.e. its (mass$)^2$ is negative implying
that it travels faster than the speed of light. The presence of such a 
particle makes the vacuum unstable, which is not acceptable in a physical
theory. 

The resolution of this tachyon problem was found in a series of
remarkable discoveries which led to the concept of
supersymmetry, a symmetry relating bosonic and fermionic
particles. 
The first discovery was the existence of a new consistent
string theory
whose spacetime dimension turns out to be 10 rather than 26. The
second discovery was that the action for this new string theory 
depends on a two-dimensional worldsheet containing
both bosonic and fermionic parameters, and the action is invariant
under a worldsheet supersymmetry which transforms 
the bosonic and fermionc parameters into each other. 
The third discovery was that, after performing a projection 
operation which removes half the particles but leaves a unitary 
S-matrix, the particle spectrum and interactions of this `superstring' theory 
are invariant under a ten-dimensional
spacetime-supersymmetry which transforms bosons into fermions. 
This projection operation removes the problematic
tachyon from the spectrum
but leaves the massless gluons, as well as an
infinite number of massive particles. Superstring theory also 
contains fermionic counterparts to the gluon (called
the gluino), as well as an infinite number of
massive fermions. 

Another particle which survives the projection operation is 
a massless spin-two particle called the graviton (as well as its
fermionic counterpart, the massless spin-3/2 particle called the
gravitino).
Although this massless spin-2 particle
comes from closed string theory (where particles
are represented by one-dimensional circles), unitarity implies
that the two ends of an open string can join to form a closed string,
so these massless spin-two particles
are produced in the scattering of gluons. 
Since the only consistent interactions of massless spin-two particles are 
gravitational interactions, string theory `predicts' the existence of gravity. 
Therefore, without prior intention, superstring theory was found to give 
a unified description of
Yang-Mills and gravitational interactions. 

Since the energy scale of gravitational interactions is much larger than
the energy scale of strong interactions, a unification of these
interactions implies that the massive particles predicted by
superstring theory contain masses of the order of the
Planck mass ( about $10^{19}$ GeV), and are therefore
unrelated to meson particles found
in experiments. So the original motivation for using string theory
as a model for strong interactions is no longer viable, assuming that
one interprets the massless spin-two particle as the graviton of general
relativity.
Instead, superstring theory can be used as a model for a unified 
theory which includes all four of the standard interactions: gravitational,
strong, weak, and electromagnetic (the last three are described by
a spontaneously broken Yang-Mills theory).

The usual obstacle to constructing a quantum unified theory (or even
a quantum theory of gravity) is that
the Einstein-Hilbert action for general relativity is non-renormalizable.
This is easily seen from the fact that the gravitational coupling
constant (Newton's constant) is dimensionful, unlike the coupling constant of 
Yang-Mills theory. So for a scattering amplitude of three gravitons
at $L$ loop-order, power counting arguments imply that the amplitude
diverges like $\L^{2L}$ where $\L$ is the cutoff. The only way
to remove this divergence is if there is some miraculous cancellation
of Feynmann diagrams.

One way to cancel divergences in Feynmann diagrams is
to introduce fermions into the theory with the same interactions
and masses as the bosons. Since internal loops of fermions contribute with an
extra minus sign as compared with internal loops of bosons, there
is a possibility of cancellations. If a theory is supersymmetrized
(i.e. fermions are introduced in such a manner that the theory
is symmetric under a transformation which exchanges the bosons and fermions),
then the above conditions are satisfied. The supersymmetrization of 
gravity is called supergravity, and for a few years, it was hoped
that such a theory might be free of non-renormalizable divergences.
However, it was later realized that even after supersymmetrizing gravity to
a theory with the maximum number of supersymmetries (which is called
N=8 supergravity), the non-renormalizable divergences are still present.

As already mentioned, the fundamental particles of superstring theory
include the graviton and the gravitino (like supergravity), but also
include 
an infinite set of massive bosons and fermions. It turns out that
after including the contributions of the infinite massive particles,
the non-renormalizable divergences in the loop amplitudes
completely cancel each other out. Although the explicit proof of
the preceding statement is rather technical, there are various `handwaving'
arguments which are convincing. One of these arguments involves the
nature of superstring interactions which are `smoother' than the
interactions of point-particles. For example, the three-point diagram
for point-particles has a vertex where the three external 
point-particles coincide. But
the three-point diagram for closed strings is like a pair of pants, where
the two cuffs and the waist are the external strings. Unlike the vertex in a
point-particle diagram, there is no singular point on a pair of pants.

So superstring theory provides a consistent theory
of quantum gravity which, unlike all other attempts, does not suffer
from non-renormalizable divergences. 
However, it requires an infinite
set of massive particles which are unobservable in any foreseeable 
experiment. In addition, the theory includes a set of massless particles
such as the gluons and gluinos
of super-Yang-Mills and also a scalar massless boson 
called the dilaton. If superstring theory really
describes nature (and is not just a model for a unified quantum theory
of gravity and Yang-Mills), these massless particles must become
the leptons, quarks, and gluons of the standard model where the masses
of the above particles come from spontaneous symmetry breaking. 
One important unsolved problem in superstring theory is that
it is very difficult to give a mass to the dilaton in a natural way,
so one needs to explain why noone has observed
massless scalars in experiments.  

Although superstring theory is the only candidate for a renormalizable
quantum theory
of gravity, only a few researchers worked in this field between
1975 (when it was realized that string theory could not serve as a model
for strong interactions) and 1985. One reason for the lack of interest was
that there appeared to be different versions of superstring theory
(called Type I, Type IIA and Type IIB), none of which resembled very closely
the structure of the standard model. 
In the Type I theory, the gauge group for super-Yang-Mills 
was thought to be
arbitrary, and in the Type IIA and Type IIB theories, the gauge group
had to be abelian. However, in 1985, it was learned that absence of 
anomalies restricted the gauge group of the Type I theory to be 
$SO(32)/Z_2$. Although this gauge group is not very interesting for
phenomenology, it was soon realized that there is another type
of superstring theory, called the `heterotic' superstring (since it 
combines features of the bosonic string and superstring), which has two
possible gauge groups: $SO(32)/Z_2$ or $E_8\times E_8$ ($E_8$ is
one of the exceptional groups). The $E_8\times E_8$
version of the heterotic superstring was very attractive for phenomenologists
since it is easy to construct grand unified theories starting from the
exceptional subgroup $E_6$. 

For this reason (and because of peer pressure), the next five years 
attracted many researchers into the field of superstring theory. However, 
it was soon clear that without understanding non-perturbative effects,
superstring theory would not be able to give explicit
predictions for a grand unified model (other than vague predictions, such
as supersymmetry at a suitably high energy scale).
The problem was that four-dimensional
physics depends crucially on the type of compactification which is used to
reduce from ten to four dimensions. Although there is a symmetry
called $T$-duality which relates some 
compactifications in superstring theory, there is a large
class of compactifications which are not related by any symmetry. 
In principle, the type of
compactification is determined dynamically, however, the selection of the
correct compactification scheme requires 
non-perturbative information. So, for this reason (and because of
problems in finding jobs), many researchers left
the field of string theory after 1989
to work in other areas such as supercollider
phenomenology.

Recently, it has been learned that many non-perturbative features of
four-dimensional
supersymmetric Yang-Mills theories can be understood without performing
explicit instanton computations. Although this had been conjectured in 1977
for N=4 super-Yang-Mills, 
the conjecture was treated skeptically until 1994 when convincing 
evidence was presented for the case of N=2 super-Yang-Mills.
One of these non-perturbative features is an `$S$-duality' symmetry which
relates the super-Yang-Mills theory at large values of the coupling constant
with a super-Yang-Mills theory at small values of the coupling constant.
For N=4 super-Yang-Mills, $S$-duality maps the theory at strong coupling
into the same theory at weak coupling, while for N=2 super-Yang-Mills,
$S$-duality maps the theory at strong coupling into a different theory
at weak coupling. 

These $S$-duality symmetries are also believed to be present in superstrings
and relate superstring theory at 
large values of the coupling constant with
a theory at small values of the coupling constant. 
$S$-duality maps the
Type IIB superstring at strong coupling into the same Type IIB superstring
at weak coupling, and maps the
Type I superstring at strong/weak coupling into the heterotic superstring
at weak/strong coupling with gauge group
$SO(32)/Z_2$. 

There is also believed to a duality symmetry
which maps the Type IIA superstring
at strong coupling into a new eleven-dimensional theory called 
$M$-theory,
and which maps the heterotic superstring
with gauge group $E_8\times E_8$ at strong coupling
into a version of $M$-theory with 
boundaries.
$M$-theory is known to contain the massless particle of
eleven-dimensional 
supergravity (which is the maximum possible dimension for
supergravity) as well as massive particles which are still not understood.
It is believed to be related to a theory constructed from two-dimensional
extended objects called membranes (as opposed to the one-dimensional 
extended objects called strings). 

So by studying the perturbative regime of superstring theory where
the coupling constant is small, one can use $S$-duality symmetry to
obtain non-perturbative information where the coupling constant is large.
Furthermore, duality symmetries relate the five different
superstring theories, suggesting that these five theories can be understood as
perturbative vacua of some unique underlying non-perturbative theory 
which would be the `Theory of Everything'.
This has attracted renewed interest in superstring theory, and there
is optimism that by studying $M$-theory, one will gain a greater understanding
of duality symmetries. However, the problem of getting explicit predictions
out of superstring theory is probably still far from being resolved. 
Although $S$-duality
symmetries may help in understanding superstring theory at very small
and very large values of the coupling constants, it is not clear if
it will be possible to extrapolate these results to the physically
interesting values
of the coupling constants which is somewhere between the two extremes. 

In section 2 of this paper, I will discuss classical relativistic
strings.
In section 3, I will show how to quantize the relativistic string
and compute the spectrum. In section 4, I shall introduce the Type IIA
and Type IIB superstrings
in the light-cone Green-Schwarz approach. In section 5, I shall discuss
compactification and T-duality. In section 6, I will describe
eleven-dimensional supergravity and give a simple argument for the
$S$-duality symmetry of the Type IIB superstring. 

\newsec{Classical Relativistic Strings}

As is well known, the action for relativistic point-particles moving 
in $D$
dimensions is given by
\eqn\one{ S=M\int_{\tau_I}^{\tau_F} d\tau L(x^\mu (\tau))=M
\int_{\tau_I}^{\tau_F} d\tau \sqrt{\p_\tau x^\mu \p_\tau x_\mu} }
and the equation of motion is $M \p_\tau 
(\p_\tau x^\mu/\sqrt{(\p_\tau x)^2})=0.$ In the above action, $\mu=0$ to
$D-1$, $M$ is
a dimensionful constant,
and $L(x)$ is defined
as the length of the path traversed by $x^\mu(\tau)$
between the times $\tau_I$ and $\tau_F$.
The momentum is defined by 
\eqn\two{ P_\mu =M{{\p L}\over{\p (\p_\tau x^\mu)}} = M {{\p_\tau x_\mu}\over
{\sqrt{(\p_\tau x)^2}}},}
so $P_\mu P^\mu = M^2$ where $M$ is identified with the mass of the particle.

The above action is invariant under reparameterizations of the worldline,
$\tau \to \tilde\tau (\tau)$, allowing the gauge choice $\p_\tau x^\mu 
\p_\tau x_\mu =1$. In this gauge, the equation of motion becomes
$M \p^2_\tau x^\mu =0$, which has the solution
$$x^\mu (\tau) = x_0^\mu +\tau P^\mu$$
where $P^\mu P_{\mu}=M^2$. 

For a relativistic one-dimensional object with the topology of a closed string
(i.e. the topology of a circle),
the obvious generalization of 
\one is 
\eqn\act {S={T\over {2\pi}}\int_{\tau_I}^{\tau_F} d\tau \oint_0^{2\pi} d\sigma
 A(x^\mu (\tau, \s))}
$$={T \over {2\pi}}
\int_{\tau_I}^{\tau_F} d\tau 
\oint_0^{2\pi} d\sigma
\sqrt{(\p_\tau x^\mu \p_\tau x_\mu)
(\p_\s x^\mu \p_\s x_\mu)-
(\p_\tau x^\mu \p_\s x_\mu)^2 }$$
where $T$ is a dimensionful constant, $\s$ is a parameter ranging from
0 to $2\pi$ which measures the position on the circle,
$\p_\s x^\mu = \p x^\mu/\p\s$, and $A(x)$ is the
area of the cylindrical 
surface traversed by $x^\mu(\tau,\s)$ between the times
$\tau_I$ and $\tau_F$. (The formula for the area is easily found by
dividing the surface into infinitesimal parallelograms whose sides are given
by $d\tau \p_\tau x^\mu $ and 
$d\s \p_\s x^\mu $.)

This action is invariant under reparameterization of the worldsurface,
$\tau \to \tilde\tau (\tau,\s)$ and 
$\s \to \tilde\s (\tau,\s)$, which allows one to choose the
gauge $\p_\tau x^\mu \p_\tau x_\mu = $
$\p_\s x^\mu \p_\s x_\mu$ and 
$\p_\tau x^\mu \p_\s x_\mu=0$. In this gauge, it is easy to show that
the equation of motion from \act is
\eqn\motion{\p_\tau^2 x^\mu =\p_\s^2 x^\mu }
and the momentum is defined by 
\eqn\mom{P_\mu = {T\over {2\pi}} {{\p A}\over{\p(\p_\tau x^\mu)}}=  
{T\over {2\pi}} \oint_0^{2\pi} d\s \p_\tau x_\mu.}
So the general solution to the equation of motion is 
\eqn\sol{
x^\mu (\tau,\s) = x_0^\mu +{1\over T}
\tau P^\mu +\sum_{N=-\infty, N\neq 0}^{\infty}
(a_N^\mu e^{iN(\tau+\s)} +\tilde a_N^\mu e^{iN(\tau-\s)}).}
The gauge-fixing conditions 
\eqn\fix{\p_\tau x^\mu \p_\tau x_\mu -
\p_\s x^\mu \p_\s x_\mu = 
\p_\tau x^\mu \p_\s x_\mu=0}
imply that two of the $D$ components of $x^\mu$ can be related to the
other $D-2$ components and that
\eqn\mass{P^\mu P_{\mu}=T^2  
\sum_{N=-\infty, N\neq 0}^{\infty} N^2
(a_N^j a_{-N}^{j} +\tilde a_N^j \tilde a_{-N}^{j})}
where $j$=1 to $D-2$. 

Since 
$P^\mu P_{\mu}$ gives the (mass$)^2$ of the string, the mass
of the string depends on the $a_N^j$ and $\tilde a_N^j$ variables,
and therefore depends on the way that the string is resonating.
Each distinct resonance of the string corresponds to a different
particle whose mass can be computed from \mass. 
Although the classical relativistic string has a continuous mass spectrum,
the spectrum will become discrete after quantization. 

Note that $T$ corresponds to the tension of the string since
\motion and \mom imply that
$\p_\tau \hat P_j = T  \p_\s^2 x^j$ where $\hat P_j$ is the momentum
density (i.e. $P_j ={1\over {2\pi}}\oint_0^{2\pi} d\s \hat P_j$).
In natural units for describing gravitational
interactions, $T$ is approximately $(10^{19} GeV)^2$.   .                

\newsec{Quantization of the Closed String}

In the previous section, it was seen that 
\eqn\dens{\hat P^\mu(\tau,\s)=T \p_\tau x^\mu = P^\mu +i T 
\sum_{N=-\infty, N\neq 0}^{\infty} N
(a_N^\mu e^{iN(\tau+\s)} +\tilde a_N^\mu e^{iN(\tau-\s)}).}
Using \sol and the canonical commutation relations
\eqn\can{ [x^\mu (\tau, \sigma), \hat P^\nu(\tau, \sigma ')]= 
i  \eta^{\mu\nu} \delta (\s - \s '),} 
one finds that $a_N^\mu$ and $\tilde a_N^\mu$ satisfy the commutation 
relations 
\eqn\com{ 
[ a_M^\mu , a_N^\nu]= 
{1\over {T N}}  \d_{M+N,0}\eta^{\mu\nu},\quad 
[\tilde a_M^\mu ,\tilde a_N^\nu]= 
{1\over {T N}}  \d_{M+N,0}\eta^{\mu\nu}.} 
As in the harmonic oscillator,
one can define a ground state $|0\>$ which is annihilated
by $a_N^{\mu}$ and $\tilde a_N^\mu$ for $N<0$.

So using \mass, the state 
\eqn\state{|\Phi\>= \prod_{j=1}^{D-2}\prod_{N> 0} (a_N^j)^{n_N^j} 
(\tilde a_N^j)^{\tilde n_N^j}|0\>} 
has
(mass$)^2$ given by the formula of \mass,
\eqn\spec{M^2 =T^2 \< \Phi | 
\sum_{N> 0}N^2
(2 a_N^j a_{-N}^j +[a_{-N}^j, a_N^j] +2\tilde a_N^j \tilde a_{-N}^{j}+ 
[\tilde a_{-N}^j,\tilde a_N^j] ) |\Phi\>.}
Plugging \state into \spec and using the commutation relations of \com,
one finds
\eqn\ans{M^2 = 2T
\sum_{N> 0}\sum_{j=1}^{D-2} 
N (n_N^j +\tilde n_N^j) + 2T \sum_{N>0} N (D-2) }
where the second term comes from normal-ordering (as in the ground-state
energy of the harmonic oscillator). 

To compute this normal-ordering
term, one
uses zeta-function regularization to remove the divergence. This means 
defining 
$\sum_{N>0} N$ as the analytic continuation as $s\to -1$ of 
$\sum_{N>0} N^{-s}$. This analytic continuation gives
$\sum_{N>0} N = -{1\over 12},$ so the normal-ordering term (which
gives $M^2$ for the ground-state) is equal to $2 T (2-D)/12$. 
So when $D>2$,
the ground-state has negative (mass$)^2$ and is tachyonic as
described in the introduction. This means that the vacuum is unstable,
implying that closed string theory is inconsistent. As shown in the
following section, this inconsistency
is not present in closed superstring theory. 

Note that
the spin-two state in closed string theory is described by 
$|\Phi^{\mu\nu}\> = (a_1^\mu \tilde a_1^\nu
+a_1^\nu \tilde a_1^\mu)|0\>$, which has $M^2= 2T (26 -D)/12$ using
the formula of \ans. So when $D=26$ this spin-two state
is massless and describes
a graviton. 

\newsec{Type II Superstrings in the Light-Cone Green-Schwarz Approach}

There are many equivalent descriptions of the Type IIA and
Type IIB superstrings, but the only
one which will be described in these notes is the 
formalism of Green and Schwarz in the light-cone gauge. 
Light-cone gauge means that the constraints of \fix have been
used to eliminate two of the spacetime variables so
one is left with the variables $x^j(\tau,\s)$ for $j=1$ to $D-2$.
Unitarity and Lorentz invariance imply that $D=10$ for the superstring,
so $j$ takes the values 1 to 8.

However, unlike the string theory of the preceding section, the Type II
superstring
also contains fermionic variables, $\t^\a (\tau,\s)$ and
$\tilde\t^A (\tau,\s)$, where $\a$ and $\ad$ are the chiral and anti-chiral
eight-dimensional
spinor representations of SO(8), $A$ is in the $\ad$ representation
for the Type IIA superstring, and $A$ is in the $\a$ representation
for the Type IIB superstring. The chiral and anti-chiral eight-dimensional
spinor representations, $\a$ and $\ad$, are
defined using SO(8) Pauli matrices, $\s^j_{\a\dot\a}$, which satisfy
the anti-commutation relations
\eqn\pauli{ \s^j_{\a\ad}\s^k_{\a\bd} +
\s^k_{\a\ad}\s^j_{\a\bd}=2\d^{jk} \d_{\ad\bd}, \quad 
\s^j_{\a\ad}\s^k_{\b\ad} +
\s^k_{\a\ad}\s^j_{\b\ad}=2\d^{jk} \d_{\a\b}}
($j=1$ to 8, $\a=1$ to 8, and $\ad=1$ to 8). 
Note that $\a$ and $\ad$ resemble the
two-component spinor representations of SO(3,1), however in the case of
SO(8), they are independently real ($(\a)^* =\a$ and $(\ad)^*=\ad$) as
opposed to the case of SO(3,1) where $(\a)^* =\ad$. Also, SO(8) spinor
indices can be raised and lowered using the 
trivial metric 
$\d^{\a\b}$ and $\d^{\ad\bd}$.

In light-cone gauge, 
the action for the Type II superstring is given by 
\eqn\acts{S = {T\over {2\pi}}\int_{\tau_I}^{\tau_F} d\tau\oint_0^{2\pi} d\s
(\p_+ x^j \p_- x^j + \t^\a \p_- \t^\a +\tilde\t^A \p_+\tilde\t^A)}
where $\p_\pm =\p_\tau \pm \p_\sigma$.

The equations of motion for $x^j$ are the same as before, and
the equations of motion for $\t^\a$ and $\tilde\t^A$ are 
$\p_- \t^\a =\p_+ \tilde\t^A=0$, which has the general solution
\eqn\gen{\t^\a (\tau,\s)= \sum_N b_N^\a e^{iN(\tau+\s)},\quad
\tilde\t^A (\tau,\s)= \sum_N \tilde b_N^A e^{iN(\tau-\s)}.}
The anti-commutation relations 
\eqn\anticom{\{\t^\a(\tau,\s), T \t^\b(\tau,\s ')\}=
\d^{\a\b} \d(\s -\s'),\quad
\{\tilde\t^A(\tau,\s), T \tilde\t^B(\tau,\s ')\}= \d^{AB} \d(\s -\s')}
imply that 
\eqn\antic{\{b^\a_M, b^\b_N\}= T^{-1}
\d^{\a\b} \d_{M+N,0}~~,
\quad
\{\tilde b^A_M, \tilde b^B_N\}= T^{-1}\d^{AB} \d_{M+N,0}.}

For the Type II superstring, the ground state $|0\>$ is defined to be
annihilated by $a_N^j$, $\tilde a_N^j$, $b_N^\a$ and $\tilde b_N^A$ for
$N<0$. 
To determine the spectrum,
one uses the
superstring version of the 
gauge-fixing constraints of
\fix which implies that
the spectrum for the superstring is given by
\eqn\smass{M^2=P^\mu P_{\mu}=T^2  
\sum_{N=-\infty, N\neq 0}^{\infty} [
N^2 (a_N^j a_{-N}^{j} +\tilde a_N^j \tilde a_{-N}^{j})
+
N (b_N^\a b_{-N}^{\a} +\tilde b_N^A \tilde b_{-N}^{A})].}

For a state
\eqn\state{|\Phi\>= \prod_{j=1}^8\prod_{N> 0} 
(a_N^j)^{n_N^j} 
(\tilde a_N^j)^{\tilde n_N^j}\prod_{\a,A=1}^8\prod_{M\geq 0}
(b_N^\a)^{m_M^\a} 
(\tilde b_M^A)^{\tilde m_M^A}
|0\>,} 
the (mass$)^2$ is  
\eqn\spec{M^2 =T^2 \< \Phi | 
\sum_{N\geq 0}[N^2
(2 a_N^j a_{-N}^j +[a_{-N}^j, a_N^j] +2\tilde a_N^j \tilde a_{-N}^{j}
+[\tilde a_{-N}^j,\tilde a_N^j])  }
$$+N( 
2 b_N^\a b_{-N}^\a -\{b_{-N}^\a, b_N^\a\} +2\tilde b_N^A \tilde b_{-N}^A
-\{\tilde b_{-N}^A,\tilde b_N^A\})] 
 |\Phi\>.$$

Since 
$$\sum_j N^2 
[ a_{-N}^j, a_N^j] +
\sum_j N^2 
[\tilde a_{-N}^j,\tilde a_N^j]= 
\sum_\a N 
\{ b_{-N}^\a, b_N^\a \}
+\sum_A N \{\tilde b_{-N}^A,\tilde b_N^A\}, $$
the normal-ordering contribution from the $a_N^j$ and $\tilde a_N^j$ modes
is precisely cancelled by the normal-ordering contribution from the 
$b_N^\a$ and $\tilde b_N^A$ modes. Therefore, the superstring ground-state
has zero mass and the excited states carry 
\eqn\sspectrum{M^2 =2 T
\sum_{N>0}~ \sum_{j,\a,A=1}^8 N(n_N^j +\tilde n_N^j +m_N^\a + \tilde m_N^A).}

Actually, there is more than one massless state of the superstring since
hitting $|0\>$ with $b_0^\a$ and/or
$\tilde b_0^A$ does not change the mass.
Since $b_0^\a$ and $\tilde b_0^A$ satisfy the same
anti-commutation relations as the SO(8) Pauli matrices in \pauli, 
the massless state is not a scalar of SO(8) but is 
actually a 256-component multiplet of SO(8). This multiplet is described
by the 
states $|0\>^{jk}$, $|0\>^{j \dot A}$, 
$|0\>^{\ad k}$, and $|0\>^{\ad \dot A}$ where $j,k$ are SO(8)
vector representations and $\dot A$ is the opposite spinor representation
of $A$. 

The action of $b_0^\a$ and $\tilde b_0^A$ on these states is defined by
\eqn\vacrep{b_0^\a |0\>^{jk}= \s_j^{\a\ad} |0\>^{\ad k},\quad
b_0^\a |0\>^{j\dot A}= \s_j^{\a\ad} |0\>^{\ad \dot A},}
$$b_0^\a |0\>^{\ad k}= \s_j^{\a\ad} |0\>^{j k},\quad
b_0^\a |0\>^{\ad\dot A}= \s_j^{\a\ad} |0\>^{j \dot A},$$
$$\tilde b_0^A |0\>^{jk}= \s_j^{A\dot A} |0\>^{j \dot A},\quad
\tilde b_0^A |0\>^{j\dot A}= \s_k^{A\dot A} |0\>^{j k}$$
$$\tilde b_0^A |0\>^{\ad k}= \s_k^{A\dot A} |0\>^{\ad \dot A},\quad
\tilde b_0^A |0\>^{\ad\dot A}= \s_k^{A\dot A} |0\>^{\ad k}.$$
Note that $b_0^\a$ and
$\tilde b_0^A$ are anti-commuting, so 
$|0\>^{jk}$ and $|0\>^{\ad \dot A}$ are bosonic states while
$|0\>^{\ad k}$ and $|0\>^{j \dot A}$ are fermionic states. 

Decomposing $|0\>^{jk}$ into its symmetric, anti-symmetric, and trace
parts, one finds a graviton $g^{jk}$, a `Kalb-Ramond' field $B^{jk}$, 
and a scalar dilaton field $\phi$.
Decomposing $|0\>^{j\dot A}$ 
and $|0\>^{\ad j}$, one finds two gravitinos, $\psi_j^\ad$ and 
$\psi_j^{\dot A}$,
and two 
dilatinos, $\chi^\a$ and $\chi^A$. 
Decomposing $|0\>^{\ad\b}$ for the Type IIA superstring into
$\s^j_{\ad\b} |0\>^{\ad\b}$ and
$\s^{jkl}_{\ad\b} |0\>^{\ad\b}$,
one finds a one-form $A^j$ and an anti-symmetric three-form $A^{jkl}$. 
Finally, decomposing
$|0\>^{\ad\bd}$ for the Type IIB superstring into
$|0\>^{\ad\ad}$,
$\s^{jk}_{\ad\bd}|0\>^{\ad\bd}$
and $\s^{jklm}_{\ad\bd}|0\>^{\ad\bd}$,
one finds a scalar $A$, an anti-symmetric two-form $A^{jk}$, and a self-dual
anti-symmetric four-form $A^{jklm}$ where self-dual means that 
$\epsilon^{j_1 ... j_8} A^{j_5 ... j_8}= 70~ A^{j_1 ... j_4}$.

These massless states of the superstring
are the same as the states of Type IIA and
Type IIB supergravity in ten dimensions. However, of course, the
superstring also includes an infinite set of massive fields which
are not present in pure supergravity theories.

\newsec{Compactification and T-Duality}

Lorentz invariance and unitarity imply that superstrings propagate
in ten spacetime dimensions, so one needs an explanation for the fact that
only four spacetime dimensions are experimentally observable. One possible
explanation is that six of the nine spatial directions are
constrained to lie on small circles of radius $R$. 

It will be shown here that string theory, unlike point-particle theory,
predicts a symmetry called $T$-duality which relates compactification
on a circle of radius $R$ to compactification on a circle of radius
$(RT)^{-1}$ where $T$ is the string tension. This means that the radius
of compactification can always be chosen larger that $T^{-\half}$,
which has important implications for gravity at the Planck scale since
$T^{-\half}$ is approximately $10^{- 32}$ cm.

First, note that 
the wave-function $e^{i P^\mu x_\mu}$ should be single-valued
when $x^9 \to x^9 +2\pi R$ if $x^9$ is a compactified direction.
So the momentum $P^9$ must be equal
to $n R^{-1}$ for some integer $n$. 

Next, note that 
$x^9 (\tau, \s+2\pi)$ must equal 
$x^9(\tau,\s) +2\pi m R$ where $m$ is an integer which counts the
number of times that the closed string winds around the compactified
direction.
This means that the solution to the equation of motion of \motion
is 
\eqn\gener{
x^9 (\tau,\s) = x_0^9 +{{n\tau}\over {TR} } + \sigma m R
 +\sum_{N=-\infty, N\neq 0}^{\infty}
(a_N^9 e^{iN(\tau+\s)} +\tilde a_N^9 e^{iN(\tau-\s)}).}
Plugging into the $M^2$ formula coming from \fix, one learns that
\eqn\plug{M_9^2 = (P_0)^2 - (P_1)^2 - ... - (P_8)^2  =
 P_\mu P^\mu + (P_9)^2 = M_{10}^2 + ({n\over R})^2}
$$=
({n \over R})^2 + 
(m T R)^2  +
2 T
\sum_{N>0}~ \sum_{j,\a,A=1}^8 N(n_N^j +\tilde n_N^j +m_N^\a + \tilde m_N^A)$$
where $M_9$ is the mass measured by a nine-dimensional observer,
$M_{10}$ is the mass measured by a ten-dimensional observer, and
the $(m T R)^2$ term comes from the $T^2 \p_\s x^\mu \p_\s x_\mu$ contribution
to $M_{10}^2$.

It is easy to see from \plug that the nine-dimensional mass spectrum
is invariant under switching $R$ with $(TR)^{-1}$ if one also switches
momentum excitations $n$ with winding-mode excitations $m$.
Note that T-duality, unlike S-duality discussed in the following secton,
does not transform the string coupling constant and can therefore
be verified perturbatively.

For the Type II superstring, T-duality states that the Type IIA superstring
compactified on a circle of radius $R$ is equivalent to the Type IIB
superstring
compactified on a circle of radius $(TR)^{-1}$. The reason Type IIA
and Type IIB switch places is that switching momentum excitations with
winding excitations is only a symmetry of the Type II superstring if
$\tilde \t^A$ switches its SO(8) chirality.

\newsec{D=11 Supergravity and $M$-Theory}

Eleven dimensions is the maximum dimension in which gravity can be
supersymmetrized in a consistent manner. The bosonic
fields of D=11 supergravity
are a graviton $\hat g_{MN}$ and an anti-symmetric
three-form $\hat A_{MNP}$ where $M=0$ to 10.
Although this supergravity
theory is not renormalizable, its classical action can
be constructed and the bosonic contribution to this action is
\eqn\elact{S_{11}= {1\over \l^2}
\int d^{11} x [\sqrt{\det \hat g} (\hat R +\hat F_{MNPQ} 
\hat F^{MNPQ})}
$$+ \epsilon^{M_1 ... M_{11}}\hat F_{M_1 ... M_4}
\hat F_{M_5 ... M_8}
\hat A_{M_9 M_{10} M_{11}} ]$$
where $\l$ is the gravitatonal coupling constant
and $\hat F_{MNPQ}= \p_{[M} \hat A_{NPQ]}$ is the 
field-strength for $\hat A_{MNP}$.

After compactification on a circle of radius $R_{10}$, these fields
reduce to the massless bosonic fields of the Type IIA superstring,
$[g_{\mu\nu},B_{\mu\nu},\phi, A_\mu,A_{\mu\nu\rho}]$ where
$\hat g_{\mu\nu} = e^{-2\phi/3} g_{\mu\nu}$, $\hat g_{10~10}=e^{4\phi/3}$,
$\hat g_{\mu~10}=e^{4\phi/3}
A_\mu$, $\hat A_{\mu\nu~10}=B_{\mu\nu}$ and
$\hat A_{\mu\nu\rho}=A_{\mu\nu\rho}$. With this identification, the
Einstein-Hilbert part of the D=11 action
${1\over \l^2}
\int d^{11} x \sqrt {\det\hat g} \hat R$ 
reduces to 
${1\over \l^2}
\int d^{10} x e^{-2\phi}\sqrt {\det g} \hat R$.  
This means that the string coupling constant can be absorbed into a
redefinition of $\phi \to \phi +log \l$. After this redefinition,
the vacuum expectation value for $e^{\phi}$ becomes $\< e^{\phi}\>=\l$.
 
Since the compactification radius is proportional to $\sqrt{\hat g_{10~10}}
=e^{2\phi/3}$, $R_{10}$ is proportional to $\l^{2/3}$. Therefore, the
Type IIA superstring at weak coupling and low energies (i.e. 
the massless sector with $\l<<1$) is
equivalent to compactification of $D=11$ supergravity on a circle
of small radius.

However, the Type IIA superstring is a renormalizable
theory, so it also makes sense at high energies. This suggests that
there is a renormalizable version of D=11 supergravity 
with massive fields which 
makes sense at high-energies. This eleven-dimensional theory is called
$M$-theory and it will now be shown how eleven-dimensional Lorentz 
invariance of $M$-theory implies a strong-weak
duality of the Type IIB superstring.

The classical
low-energy effective action for the Type IIB superstring (i.e. the
classical action for ten-dimensional Type IIB supergravity) is known
to contain a classical symmetry called $S$-duality which transforms
the massless bosonic Type IIB fields as:
\eqn\trans{\rho \to\ {{a\rho +b}\over {c\rho +d}},\quad
B_{\mu\nu}\to a B_{\mu\nu} + b A_{\mu\nu},\quad
A_{\mu\nu}\to c B_{\mu\nu} + d A_{\mu\nu}}
$$g_{\mu\nu} \to g_{\mu\nu}, \quad
A_{\mu\nu\rho\sigma}\to A_{\mu\nu\rho\sigma}$$
where $\rho= A+i e^{-\phi}$ and $a,b,c,d$ are integers satisfying
$ad-bc=1$. 

When $a=d= A(x)=0$ and $b=-c=1$, this $S$-duality symmetry
transforms $e^{-\phi}$ to $e^{\phi}$, and since $\<e^{\phi}\>=\l$, it
takes $\l \to \l^{-1}$ which switches strong and weak coupling. This
strong-weak duality symmetry of the classical action can be proven to
be a symmetry of the full quantum Type IIB superstring action using the
following argument:

Suppose one compactifies two of the eleven dimensions of $M$-theory 
on small circles of radius $R_1$ and $R_{2}$. If $R_{1}$ is
identified with the eleventh dimension, this corresponds to a 
Type IIA superstring with $\l=(R_{1}/R_2)^{3/2}$ which is compactified
on a small circle of radius $R_2$. By $T$-duality, this corresponds to a
Type IIB superstring with 
$\l=(R_{1}/R_2)^{3/2}$ which is compactified
on a large circle of radius $(T R_2)^{-1}$. 

But by eleven-dimensional Lorentz covariance of $M$-theory, one could
also have identified $R_2$ with the eleventh dimension.
In this case, the $M$-theory compactification
corresponds to a 
Type IIA superstring with $\l'=(R_{2}/R_1)^{3/2}$ which is compactified
on a small circle of radius $R_1$. By $T$-duality, this corresponds to a
Type IIB superstring 
with $\l'=(R_{2}/R_1)^{3/2}$ which is compactified
on a large circle of radius $(T R_1)^{-1}$. 

If $R_1\to 0$ and $R_2\to 0$ with $R_1/R_2 =C$ held fixed, the two
Type IIB superstrings become uncompactified but their coupling
constants remain fixed at the value $\l={\l'}^{-1}=C^{3/2}$.
Since these two descriptions come from the same compactification of
$M$-theory, the uncompactified Type IIB superstring is invariant
under an $S$-duality symmetry which exchanges $\l$ and $\l^{-1}$, and
therefore exchanges strong and weak couplings..

{\bf Acknowledgements:} I would like to thank
the organizers of the conference in Vitoria for a very enjoyable
week of lectures. This
work was financially supported by 
FAPESP grant number 96/05524-0.

\end